\documentstyle[aps,multicol]{revtex}
\draft  
\begin{document}

\title{Quantization of Gauge Theory for Gauge Dependent Operators}
\author{Xiang-Song Chen$^{a,b\thanks{present address}}$, 
	Wei-Min Sun$^a$, Fan Wang$^a$, Amand Faessler$^b$}
\address{$^a$Department of Physics and Center for Theoretical Physics, 
        	Nanjing University, Nanjing 210093, China\\
	$^b$Institute for Theoretical Physics, University of Tuebingen,
	        Auf der Morgenstelle 14, D-72076 Tuebingen, Germany}
\date{August 1999}
\maketitle

\begin{abstract}
Based on a canonically derived path integral formalism, we demonstrate 
that the perturbative calculation of the matrix element for gauge dependent 
operators has crucial difference
from that for gauge invariant ones. For a gauge dependent 
operator ${\cal O}(\phi)$ what appears in the Feynman diagrams is not 
${\cal O} (\phi)$ 
itself, but the gauge-transformed one ${\cal O}(^\omega \phi)$, where $\omega$
characterizes the specific gauge transformation which brings any field variable
into the particular gauge 
which we have adopted to quantize the gauge theory using the canonical method. 
The study of the matrix element of 
gauge dependent operators also reveals that the formal path integral formalism 
for gauge theory is not always reliable. 

\pacs{PACS number: 11.15.-q}
\end{abstract}

\begin{multicols}{2}
Ever since the emergence of nucleon spin problem, the matrix element of gauge
dependent operators has aroused great interest \cite{spin}. This is because the
conventional understanding of gluon spin, gluon and quark orbital angular
momentum all corresponds to gauge dependent operators. When studying these
operators, people have naturally adopted the same calculation scheme as for
gauge invariant ones, however some quite contradictory results were
obtained \cite{Hood}. 
In this paper, by quantizing the gauge theory from the
very beginning, we demonstrate that the perturbative calculation rules for 
gauge dependent operators have nevertheless critical difference from those for
gauge invariant ones. By doing so we also reveal that the highly formal 
path-integral formalism for gauge theory is not always reliable. 

The theory we start with is the classical SU(N) or U(1) gauge theory. We adopt
the well-defined canonical quantization approach to quantize it. Due to gauge
freedom, we must specialize a gauge at first. Here we choose the most
convenient axial gauge
\begin{equation}
   n\cdot A =0, \label{gauge}
\end{equation} 
where $A$ is the gauge potential, $n$ is an arbitrary but fixed constant 
four-vector. Now we can apply the standard canonical quantization procedure and
obtain the commutation relations and the quantum Hamiltonian. But instead of 
going further to use the operator method to derive the Feynman rules, we use 
the so obtained quantum Hamiltonian to derive the path-integral expression for 
the vacuum expectation value of an arbitrary operator ${\cal O}(\phi)$, where 
$\phi$ denotes collectively the gauge and matter fields. After the standard 
procedure which can be found in textbooks \cite{Wein}, 
we can write up to an irrelevant normalization factor 
\begin{equation}
   Z({\cal O})\equiv \left\langle {\cal O} \right\rangle _{\rm vac} =
	\int[D\phi] e^{iS}{\cal O}(\phi)\delta(n\cdot A).   	\label{nonp}
\end{equation}	
The notations are standard. This expression is rigorous and applies to both
Abelian and non-Abelian gauge theory \cite{note}. 
Moreover, we have no requirement for the
operator ${\cal O}$. It can be either gauge dependent or gauge invariant. 
However, Eq. (\ref{nonp}) is non-perturbative and not 
yet perturbatively applicable. To derive the rules for perturbative 
calculation, we
use the Faddeev-Popov trick by multiplying Eq. (\ref{nonp}) with the identities
\begin{mathletters}%
\begin{eqnarray}
   1&=&\Delta_F(A,\chi) \int [D\omega] \delta (F(^\omega A) -\chi), 
   	\label{det}\\
   1&=&\int [D\chi] G[\chi], \label{G}
\end{eqnarray}
\end{mathletters}%
where $F$ and $G$ are two arbitrary functions (or in general functionals),
$[D\omega]$ is the invariant
measure of the gauge group, $^\omega A$ is the result of $A$ after a gauge
transformation characterized by $\omega$.
$F$ is usually called the gauge
fixing function, for example we can take our original axial gauge form 
$F(A) =n\cdot A$.  
Accidentally, it is apparent that by this method we can transit to an arbitrary
gauge by choosing arbitrary $F$ and $G$ 
(this point will be explained further later).
        
Multiplying (\ref{nonp}) with (\ref{det}) and (\ref{G}), we get
\begin{eqnarray}
        Z({\cal O})=\int &&[D\phi][D\omega] [D\chi]
	  e^{iS} {\cal O} \delta (n\cdot A) \nonumber \\
          && \times \Delta_F(A,\chi) \delta (F(^{\omega} A) -\chi) 
         G[\chi]. \label{step1}
\end{eqnarray}
Now make the transformation 
\begin{equation}
        \phi \rightarrow ^{\omega{^-1}} \phi , \label{trans}
\end{equation}	
note that $[D\phi]$, $S$, and $\Delta_F(A,\chi)$ are gauge invariant, and 
integrate over $\chi$, we get
\begin{eqnarray}
        Z({\cal O})&=&\int [D\phi] e^{iS} \Delta_F(A,F(A) G(F(A)) \nonumber \\
         && \times \int [D\omega] (^{\omega^{-1}} {\cal O}) 
	 	\delta (n \cdot ^{\omega^{-1}} A) 
	 \label{step2}
\end{eqnarray}	

The second line in Eq. (\ref{step2}) gives up to an irrelevant constant factor
\cite{note1}
\begin{equation}
        ^{\omega_0} {\cal O}, \label{omega0}
\end{equation}
where $\omega_0$ is the specific function of $A$ which satisfies 
\begin{equation}
        n \cdot ^{\omega_0} A =0. \label{def}
\end{equation}	
So we obtain the aimed expression 
\begin{equation}
   Z({\cal O})=\int [D\phi] e^{iS} (^{\omega_0} {\cal O}) 
   	\Delta_F(A,F(A)) G(F(A)). \label{aim}
\end{equation}
By choosing a Gaussian type of $G$, and introduce the ghost field to treat the
Faddeev-Popov determinant $\Delta_F$, one can straightforwardly derive Feynman 
rules from Eq. (\ref{aim}).

We note very importantly that what appears in this final expression is 
$^{\omega_0} {\cal O}$ instead of ${\cal O}$ itself. This will give critical
difference for gauge invariant and non-invariant operators. 

First we look at the gauge invariant operator: $^{\omega_0} {\cal O}$ equals 
${\cal O}$. Now Eq. (\ref{aim}) becomes
\begin{equation}
   Z({\cal O})=\int D\phi e^{iS} {\cal O} \Delta_F(A,F(A)) G(F(A)). 
     \label{aim'}
\end{equation}
So we see that starting from a canonical quantization in the specific axial 
gauge, we
finally arrive at an expression independent of the original gauge condition!
For instance the gauge $A_2 =0$ or $A_3=0$ lead to the same Eq. (\ref{aim'}).
This states that the Green's function of gauge invariant operators are gauge
independent.

For gauge dependent operators, however, $^{\omega_0} {\cal O}$ does not equal 
${\cal O}$. So the perturbative calculation of the 
Green's function for gauge dependent operators is not a
straightforward generalization of that for gauge invariant ones. {\em Instead, 
we encounter an additional complication of always having to apply a gauge 
transformation to the operator before inserting it into the Feynman diagrams.}

It should be remarked that in Eq. (\ref{aim}) only the operator ${\cal O}$ 
suffers 
such a gauge transformation, the other field variables such as those in the 
action remain the ordinary form.
Accordingly, in the perturbative calculations, only the external vertex  
corresponding to ${\cal O}$ should be the gauge transformed one, while the 
Feynman
rules for the internal vertices and propagators generated from the expansion of 
the effective action are still the ordinary ones.

We note that in Eq. (\ref{aim}) $\omega_0$ is universally given by
Eq. (\ref{def}). However, the form of $F$ and $G$ can be arbitrary, i.e. the
Feynman rules can be arbitrary but they give the same final results. One may
wonder why Eq. (\ref{aim}) takes such an apparently peculiar form, especially 
why must we apply a transformation to ${\cal O}$ and why
$\omega_0$ still takes the form in Eq. (\ref{def}) even after we have 
``transitted to another gauge'' by choosing an $F(A)$ different from 
$n \cdot A$.  We explain below that this actually has deeper physical reasons 
behind.

Due to the extra degree of gauge freedom, we have to fix a gauge to do
quantization. And such a gauge choice will naturally leave ``traces'' 
on the considered operator ${\cal O}$ and also on the subsequent formalism, such
as the to-be-derived quantum Hamiltonian and Feynman rules.  
Indeed, different gauges lead to different form of quantum
Hamiltonian. Or to say, the form of quantum Hamiltonian is gauge dependent. 
Furthermore, if we stick to the operator method to calculate the Green's
functions, the original gauge choice will go throughout to manifest in the 
Feynman rules. 
However the gauge invariance principle commands that we must in the end show 
all the physical observables to be independent of the original gauge we chose 
in quantization. 
Within the canonical operator method, such a demonstration is essentially a
direct proof of the equivalence of Feynman rules derived in different gauges, 
and will necessarily 
involve us in a detailed analysis of the Feynman diagrams. This was done 
for QED by Feynman \cite{Feyn}
and for Yang-Mills theory by Cheng and Tsai \cite{Cheng}.

In contrast, instead of demonstrating gauge invariance at the level of Feynman
diagrams, the advantage of path-integral formalism is that we can already get
rid of the original gauge condition at an intermediate stage {\em before} 
deriving the
Feynman rules. (To be equivalent to the canonical formalism, our
path-integral expression is derived by the operator method using the
canonically-derived quantum Hamiltonian.) The crucial step is that we 
can obtain Eq. (\ref{nonp}), in
which what appears is the original {\em gauge invariant} 
classical action $S$. (This can be done at least in the axial gauge, in
which the quantum Hamiltonian takes a relatively simpler form, than, say, in the
Coulomb gauge.) Now, the subsequent procedure in 
Eqs. (\ref{step1})-(\ref{step2})
is essentially converting $n\cdot A$ into $n\cdot A'=n\cdot ^\omega A
=(F(A)-\chi)$
by coordinate transformation, then integrating over $\chi$ with a 
weight-functional $G[\chi ]$. The $\Delta_F(A,\chi )$ is nothing but
the Jacobian of this coordinate transformation, which due to the compactness of
the gauge group can be chosen
gauge invariant. This step is mathematically trivial, and by
it alone we are still in the original axial gauge, but now expressed as 
$n\cdot A'=0$. However, since $[D^\omega \phi]$, 
$\Delta_F(^\omega A,\chi )$, and critically the action $S(^\omega \phi)$ are 
gauge invariant, we can changed their $^\omega \phi$ back to $\phi$. 
{\em Now} we are in 
{\em another} arbitrarily chosen gauge $(F(A)-\chi =0)$, 
and can commit an arbitrary weight 
$G[\chi ]$ to each $\chi $. Therefore, the freedom to choose arbitrary $F$ and
$G$ is a consequence of the gauge invariance of $S$ in Eq. (\ref{nonp}). Namely,
despite the appearance of an explicit $\delta (n\cdot A)$, the original gauge
condition $n\cdot A$ has essentially disappeared out in Eq. (\ref{nonp}) 
(apart from
${\cal O}(\phi)$), because we can convert $n\cdot A$ into an arbitrary form by
changing $A$ to $^\omega A$ but leave the $A$ elsewhere unaltered. Therefore 
later on when we convert $G$ and $\Delta_F$ into an effective action and expand
it to derive Feynman rules, we already know for sure that these Feynman rules
are equivalent to each other, without having to study the details of 
Feynman diagrams. 
This is the convenient path-integral method of demonstration that gauge theory
still reserves gauge invariance after quantization. Namely, the ``trace'' of the
original gauge choice has disappeared out in the final Feynman rules. 

However, the ``trace'' left on the considered  
operator ${\cal O}$ by the original gauge condition will not disappear, 
if ${\cal O}$ is not gauge invariant. And such a ``trace'' should
be ``visible'' no matter whatever tricks we play. Indeed, in the original
expression Eq. (\ref{nonp}), the gauge condition is enforced by the delta
function. And after we adopted the Faddeev-Popov trick to transit to an 
arbitrary gauge fixing form $G(F(A))$, this $G(F(A))$ 
no longer enforce the original gauge condition
on ${\cal O}$ \cite{note2}, however we get a gauge transformation on 
${\cal O}$ characterized by $\omega_0$.  
In fact, by the definition in Eq. (\ref{def}), 
the role of $\omega_0$ on ${\cal O}$ is nothing but bringing it to the initial 
axial gauge in which we start the canonical quantization. 

In one word, once we start with the expression Eq. (\ref{nonp}), which is
obtained by canonical quantization in the axial gauge, the subsequent
calculation will exhibit such a gauge condition anyhow, unless the considered 
operator ${\cal O}$ is gauge invariant. And no matter what kind of gauge fixing
functions $F$ and $G$ we choose to derive the Feynman rules, the final matrix
element we obtain should be the same (as long as the Faddeev-Popov trick is 
justified) and should be regarded as the result in the {\em original} 
axial gauge; the freedom of choosing
Feynman rules for calculation is simply because the quantized gauge theory still
reserves gauge invariance, as was demonstrated by the path-integral formalism. 
If we want to study what the matrix element of a gauge dependent operator would
be in another gauge, we have to begin the initial canonical quantization in
that gauge (which might not necessarily lead to a simple form like Eq.
(\ref{nonp})). Merely adopting the Faddeev-Popov trick to shift the form of $F$ 
and $G$ {\em only} changes the gauge for Feynman rules (which are equivalent 
in different gauges), but not the gauge for the {\em whole} matrix element. 

For example, the conventional understanding of quark or electron orbital
angular momentum corresponds to the gauge dependent operator 
\begin{equation}
 {\vec L}_q=\int d^3x\psi ^{\dagger }\left( {\vec x}\times 
    \frac 1i{\vec \partial }\right) \psi .
\end{equation}   
Whether ${\vec L}_q$ gives gauge invariant matrix element in a nucleon helicity
eigenstate will determine 
whether quark's orbital contribution to nucleon spin is meaningful. According to
our above 
discussion, we cannot calculate ${\vec L}_q$'s matrix element as we do for
the gauge invariant operators such as quark spin. Actually we don't really know
at all
the expression of ${\vec L}_q$'s matrix element in an arbitrary gauge except for
the gauges in which we know how to do canonical quantization. 

In the usual discussion of gauge theory with path-integral approach, people 
sometimes simply ignore the gauge freedom and write down a formal expression 
as if all field variables were independent: 
\begin{equation}
   Z({\cal O})= \int[D\phi] e^{iS}{\cal O}(\phi).   	\label{formal}
\end{equation}	
Then by following the same steps as from Eqs. (\ref{step1}) to (\ref{step2}), 
we can get 
\begin{eqnarray}
        Z({\cal O})&=&\int [D\phi] e^{iS} \Delta_F(A,F(A) G(F(A)) \nonumber \\
         && \times \int [D\omega] (^{\omega^{-1}} {\cal O}). 
	  \label{formal'}
\end{eqnarray}	
Different from Eq. (\ref{step2}), we don't have delta function in the second
line to pick up a specific $\omega$ as a function of $A$. 
The $\omega$ here is purely an integration variable independent of the field
variable. If ${\cal O}$ is gauge invariant and hence 
$^{\omega^{-1}}{\cal O}={\cal O}$, 
the infinite integration $\int [D\omega]$ factorizes out as an irrelevant
normalization factor, 
so we get the same perturbatively applicable expression as what we
obtained by starting with strict canonical quantization,
\begin{equation}
  Z({\cal O})=\int [D\phi] e^{iS} {\cal O} \Delta_F(A,F(A) G(F(A)) 
  \label{formal''}.
\end{equation}
Since Eq. (\ref{formal''}) can be derived from the same Eq. (\ref{formal}) 
for arbitrary $F$ and $G$, this formally states that $Z({\cal O})$ is 
independent of $F$ and $G$ up to an irrelevant constant factor, or to say, the
Feynman rules in different gauges are equivalent.  
This is again the same conclusion as we discovered by starting with 
well-defined canonical method.

However, we remark that such a formal procedure of starting with the ill-defined
infinite expression Eq. (\ref{formal}) is not always reliable. Such 
unreliability simply often does not manifest for gauge invariant operator, but 
can be seen by studying gauge dependent operators. Generally speaking, 
when ${\cal O}$ is gauge dependent 
we are unable to arrive at the perturbatively applicable expression Eq.
(\ref{formal''}) from Eq. (\ref{formal}). But there are some exceptions. 
For instance we take the product of electron field at different points:
\begin{equation}
{\cal O}=\psi ^\dagger (x)\psi (y). \label{psi}
\end{equation}
$^\omega {\cal O}$ will be
\begin{equation}
 ^\omega {\cal O} =e^{i\omega (x)-i\omega (y)}\psi ^\dagger (x)\psi (y). 
	\label{psi'}
\end{equation}
Therefore when $x\neq y$, $\psi ^\dagger (x)\psi (y)$ is gauge dependent.
However, we see in Eq. (\ref{formal'}) that the gauge transformed phase factor 
$e^{i\omega (x)-i\omega (y)}$ can be absorbed into the irrelevant integration 
over $\omega$. Therefore we can obtain the same expression for 
$Z(\psi ^\dagger (x)\psi (y))$ as for gauge invariant operators, and would
conclude that $Z(\psi ^\dagger (x)\psi (y))$ can be calculated with the usual
Feynman rules in an arbitrary gauge and is gauge independent! This
however contradicts the results obtained by starting with the well-defined
canonical quantization. And in fact, we known that the Fermion two point 
function is gauge dependent. 
A toy model demonstration of how such error 
arises can be found in Ref. \cite{warn}.

By some more delicate tricks, the authors in Ref. \cite{inva} adopted the above
formal path-integral formalism to demonstrate that if we use the same
calculation rules as for gauge invariant operators, the gauge dependent quark 
orbital angular momentum operator will give gauge independent matrix element 
for a nucleon helicity state. This however is refuted by explicit 1-loop 
calculations \cite{loop}.

Finally, we mention that Cheng and Tsai \cite{Cheng1} has ever pointed out
that the Feynman rules derived via the Fadeev-Popov approach often 
exhibit singularities (for example in the Coulomb and axial gauges). Such 
singularities must be dealt with care, otherwise at the two-loop level of 
non-Abelian gauge theory the Feynman rules in the Coulomb, covariant, and axial
gauges derived via the Fadeev-Popov approach will be inconsistent with each 
other \cite{Cheng1}.

In summary, we demonstrated that the 
calculation of the matrix element of gauge dependent operators is
not a straightforward generalization of that for gauge invariant ones, and must be
carefully derived from the 
very beginning of gauge theory quantization. We must distinguish the effects of 
initial gauge choice (in quantization) on
the Feynman rules and on the studied operator or matrix element. 
Feynman rules in different gauge are demonstrated to be equivalent but the 
gauge condition on the studied operator might bring non-trivial difference. 
The formal path-integral formalism
of starting with an ill-defined infinite expression, which often gives
correct results for gauge invariant operators, is however not always reliable,
as can be seen by studying the matrix element of gauge dependent operators. 

{\it Note added:} We call the interested readers' attention to a most recent
paper by Hoodbhoy and Ji \cite{Hood1}. We got quite some inspiration from 
\cite{Hood1} in developing this paper, however our opinions towards gauge
theory quantization are rather different from that in \cite{Hood1}.

\acknowledgements
One of the authors (XSC) would like to thank 
the Institute for Theoretical Physics, University of 
Tuebingen for hospitality. This work was supported by the DAAD and 
the China National Natural Science Foundation under grant No. 19675018.

\end{multicols}


\begin{references}
\bibitem{spin} For a review of the spin problem and related topics, see, e.g., 
	H.Y. Cheng, Int. J. Mod. Phys. A {\bf 11}, 5109 (1996).

\bibitem{Hood} P. Hoodbhoy, X. Ji, and W. Lu, Phys. Rev. D {\bf 59}, 074010
	(1999); and references therein. 

\bibitem{Wein} See, e.g., S. Weinberg, {\it The Quantum Theory of Fields}:
	Sections 9.6, 15.4 
	(Cambridge University Press, New York, 1995 (Vol. I), 1996 (Vol. II)).

\bibitem{note} In the simple case of QED, quantization in the Coulomb gauge can
	also lead to such a compact form. See section 9.6 of Ref. \cite{Wein}.

\bibitem{note1} This is the advantage of the axial gauge. For a general $F(A)$
	the integration of $\delta (F(^\omega A))$ over $\omega$ usually 
	acquires an $A$-dependent determinant factor, except in the U(1) case.

\bibitem{Feyn} R.P. Feynman, Phys. Rev. {\bf 101}, 769 (1949): Section 8.

\bibitem{Cheng} H. Cheng and E.C. Tsai, Chin. J. Phys. {\bf 25}, 95 (1987).

\bibitem{note2} It should be reminded that in the generally called axial gauge 
	in perturbative calculations: $F(A)=n\cdot A$ and 
	$G[\chi] =\exp (-\frac {i}{2\lambda} \int d^4x \chi^2 (x))$, 
	$n\cdot A$ is {\em not} forced to zero before we take the limit 
	$\lambda \rightarrow 0$. 
	So we must {\em still} use the gauge transformed operator 
	$^{\omega_0} {\cal O}$ in the process of calculation, 
	and take the limit $\lambda \rightarrow 0$ in the end. This does not
	necessarily give the same rusult as using ${\cal O}$ from the beginning.
	
\bibitem{warn} X.S. Chen, W.M. Sun and F. Wang, J. Phys. G (to be published); 
	the formalism in this paper and in Ref. \cite{inva}
	is taken from Ref. \cite{Wein}, it is essentially a reverse of the 
	procedure from Eq. (\ref{formal}) to Eq. (\ref{formal''}).

\bibitem{inva} X.S. Chen and F. Wang, hep-ph/9802346.

\bibitem{loop} The calculations in Ref. \cite{Hood} are for colored quark and 
	do not meet the requirement in Ref. \cite{inva}, however the 
	calculations in QED (which meet the requirement in Ref. \cite{inva}) 
	are analogous and give similar gauge dependent results 
	(X.S. Chen and F. Wang, unpublished).

\bibitem{Cheng1} H. Cheng and E.C. Tsai, Phys. Rev. Lett. {\bf 57}, 511 (1986).

\bibitem{Hood1} P. Hoodbhoy and X. Ji, hep-ph/9908275. 
\end{references}
\end{document}